\documentclass[twocolumn,aps,prl,superscriptaddress]{revtex4-1}

\usepackage{amsmath,amssymb, amstext}
\usepackage{graphicx}
\usepackage{textcomp}
\usepackage{hyperref}
\usepackage[Symbolsmallscale]{upgreek}

\usepackage[
    type={CC},
    modifier={by-nc-nd},
    version={4.0},
    imagewidth=3cm,
]{doclicense}
\begin{document}

\title{Volumetric heating of nanowire arrays to keV temperatures using kilojoule-scale petawatt laser interactions}

\author{M. P. Hill}
\email[ ]{matthew.p.hill@awe.co.uk}
\affiliation{Material Physics Group, AWE Plc, Aldermaston, RG7 4PR, UK }
\author{O. Humphries}
\author{R. Royle}
\affiliation{Clarendon Laboratory, Department of Physics, University of Oxford, OX1 3PU, UK}
\author{B. Williams}
\author{M. G. Ramsay}
\affiliation{Material Physics Group, AWE Plc, Aldermaston, RG7 4PR, UK }
\author{A. Miscampbell}
\affiliation{Clarendon Laboratory, Department of Physics, University of Oxford, OX1 3PU, UK}
\author{P. Allan}
\author{C. R. D. Brown}
\author{L. M. R. Hobbs}
\author{S. F. James}
\author{D. J. Hoarty}
\affiliation{Material Physics Group, AWE Plc, Aldermaston, RG7 4PR, UK }
\author{R. S. Marjoribanks}
\affiliation{Department of Physics, University of Toronto, M5S 1A7, Canada}
\author{J. Park}
\author{R. A. London}
\author{R. Tommasini}
\affiliation{Lawrence Livermore National Laboratory, Livermore, CA 94550, USA}
\author{A. Pukhov}
\affiliation{Heinrich Heine University of D\"{u}sseldorf, 40225 D\"{u}sseldorf, Germany}
\author{C. Bargsten}
\author{R. Hollinger}
\author{V. N. Shlyaptsev}
\author{M. G. Capeluto}
\author{J. J. Rocca}
\affiliation{Colorado State University, Fort Collins, CO 80523, USA}
\author{S. M. Vinko}
\affiliation{Clarendon Laboratory, Department of Physics, University of Oxford, OX1 3PU, UK}

\begin{abstract}
We present picosecond-resolution streaked K-shell spectra from \mbox{400 nm}-diameter nickel nanowire arrays, demonstrating the ability to generate large volumes of high energy density plasma when combined with the longer pulses typical of the largest short pulse lasers. After irradiating the wire array with \mbox{100 J}, \mbox{600 fs} ultra-high-contrast laser pulses focussed to \mbox{$>10^{20}$ W/cm$^{2}$} at the Orion laser facility, we combine atomic kinetics modeling of the streaked spectra with 2D collisional particle-in-cell simulations to describe the evolution of material conditions within these samples for the first time. We observe a three-fold enhancement of helium-like emission compared to a flat foil in a near-solid-density plasma sustaining keV temperatures for tens of picoseconds, the result of strong electric return currents heating the wires and causing them to explode and collide.
\doclicenseThis
\end{abstract}

\date{\today}

\maketitle
Nanostructured surfaces irradiated by relativistically-intense short pulse lasers present new opportunities to create ultra-high energy density systems of significant volume at near solid density, as well as high-fluence X-ray, charged particle and neutron sources. When irradiated with ultrashort laser pulses $<\mbox{100 fs}$ at intensities \mbox{$<10^{19}$ W/cm$^{2}$} nanostructured samples such as nanowire arrays have been shown to absorb almost all of the energy of the incident pulse \cite{kahaly2008near,cao2010enhanced,cerchez2018enhanced,fedeli2018ultra,Purvis2013}, volumetrically heat to a depth of several microns \cite{Bargsten2017} and exhibit significantly enhanced emission of soft X-rays compared to flat foils \cite{murnane1993efficient,gordon1994x,Kulcsar2000}. In some cases this can be as much as a 50-100$\times$ enhancement in \mbox{1-10 keV} photon yield, with laser-to-X-ray conversion efficiencies approaching 20\% \cite{Purvis2013,hollinger2017efficient}. Nanowires have been proposed as means to control the divergence of fast electrons \cite{zhao2010acceleration} and enhance or tailor the spectrum of sheath-field-accelerated MeV ions \cite{bagchi2007hot,bagchi2008hotter,Klimo2011short,sedov2019features}. As a platform for material science in extreme conditions, predictions of terabar pressures and energy densities of tens of GJ/cm$^{3}$ have been made for nanowires irradiated above \mbox{10$^{22}$ W/cm$^{2}$} \cite{Bargsten2017}, potentially creating matter in the relatively unexplored ultra-high energy density (UHED) regime.

\begin{figure}
\centering
\includegraphics[width=\linewidth]{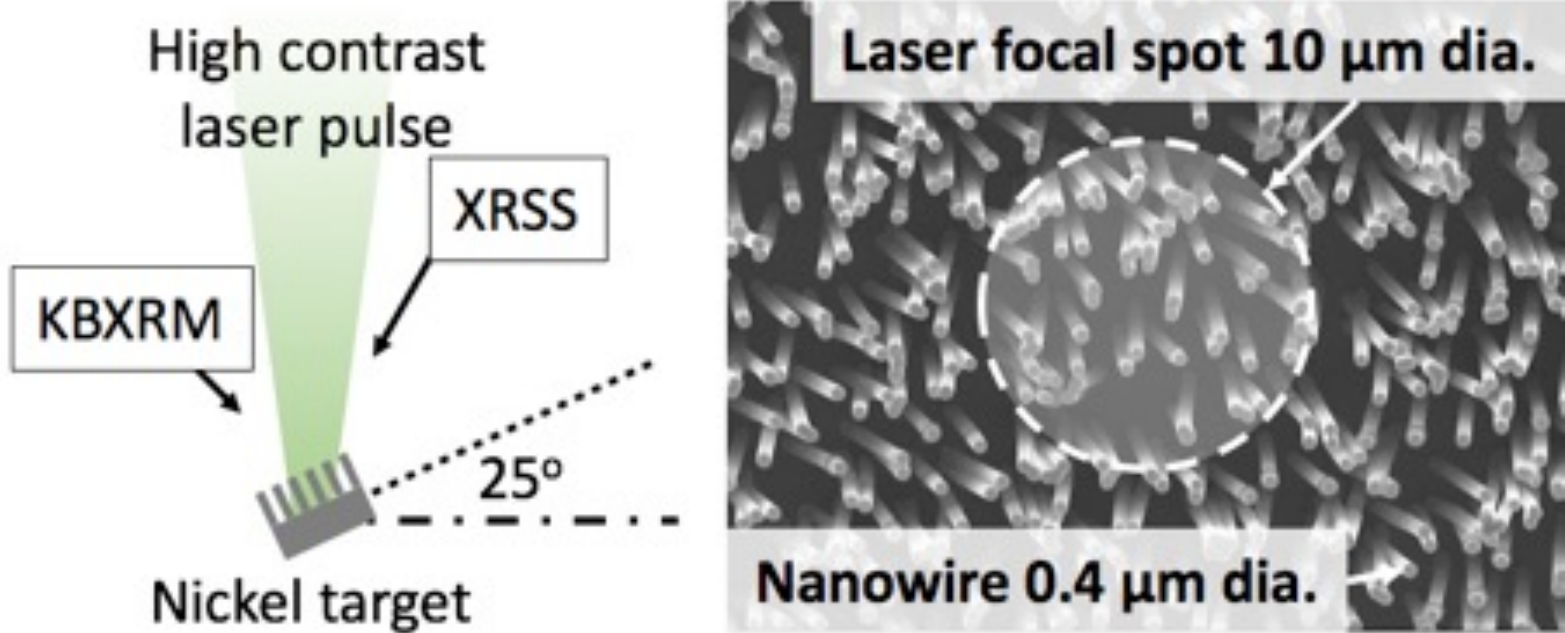}
\caption{Schematic of the experimental layout showing the positioning of the primary diagnostics, the X-ray Streaked Spectrometer (XRSS) and the Kirkpatrick-Baez X-ray microscope (KBXRM) to monitor the quality of the focal spot. A SEM image of the nickel nanowire target is also shown.}
\label{fig:one}
\end{figure}

\begin{figure}
\centering
\includegraphics[width=\linewidth]{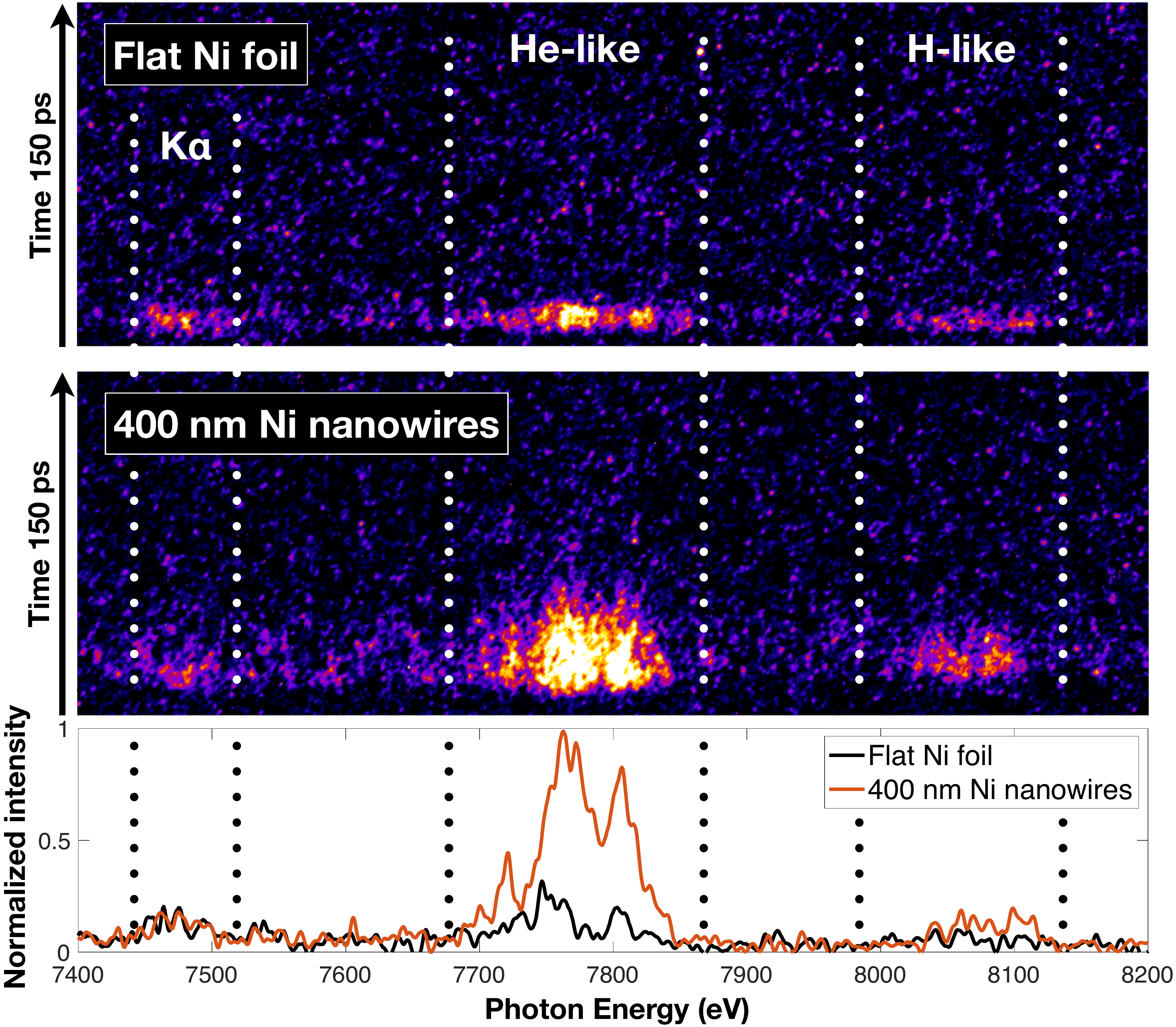}
\caption{Streaked K-shell spectra from flat and nanostructured nickel showing significant enhancement of He-like and H-like emission lines. K$_\upalpha$ emission is largely unchanged, as shown in the temporally-integrated plot below the streaked data.}
\label{fig:two}
\end{figure}

To date, experiments with nanostructured targets have been performed primarily using small-scale, joule-class laser systems. However, some of  the most promising breakthrough applications, i.e., bright X-ray and neutron source development and the generation of UHED conditions, depend to a large extent on our ability to scale the observed interaction mechanisms to much larger, more energetic laser systems. To investigate laser-nanowire interactions at high pulse energies on the picosecond timescale typical of large short pulse lasers \cite{danson2015petawatt} we have utilized the AWE Orion laser facility, where an exceptionally high contrast frequency-doubled short pulse system is capable of delivering up to \mbox{200 J} of \mbox{527 nm} light to target in a \mbox{600~fs} pulse while maintaining $10^{18}$ intensity contrast up to \mbox{100 ps} before the main pulse \cite{Hillier2014}. This contrast is necessary to avoid destroying the highly absorptive nanowire structures before the main pulse arrives \cite{cristoforetti2014investigation}. At $10^{18}$ contrast, even with peak intensities above $10^{20}\mbox{ W/cm}^{2}$, the nanosecond-scale pre-pulse does not destroy the nickel wires before the main pulse arrives, providing an excellent platform for testing model predictions. A schematic of the experimental layout and an SEM image of the nanowire sample are shown in Figure \ref{fig:one}. 

Time-resolved K-shell X-ray emission spectra between \mbox{7-9 keV} were captured for both nickel nanowire and flat targets, at \mbox{2 ps}, \mbox{$E/\Delta E>500$} resolution using a curved germanium crystal spectrometer coupled to an ultrafast X-ray streak camera, together forming the X-ray Streaked Spectrometer diagnostic (XRSS). The time-resolved experimental results are shown in Fig.~\ref{fig:two}. The wires were \mbox{400 nm} in diameter with a 15\% fill fraction and \mbox{12 \textmu m} wire length. These were irradiated with a \mbox{100 J}, \mbox{600 fs} pulse focussed to a \mbox{10 \textmu m} FWHM spot, giving an on-target intensity of $\sim10^{20}\mbox{ W/cm}^{2}$ (corresponding to a normalized vector potential $a_0\sim 6$). The quality of the focal spot was monitored by recording the time-integrated X-ray emission at \mbox{5 \textmu m}-resolution with a Kirkpatrick-Baez X-ray microscope (KBXRM). Previous work has suggested that the brightest X-ray emission can be expected along the wire axis \cite{marjoribanks2008theory,hollinger2017efficient}, in this case normal to the target surface. For this experiment the viewing angle of the spectrometer was relatively shallow (55$^\circ$ from target-normal) to prevent equipment damage, so the recorded spectra likely represent a \emph{lower bound} on the true brightness of the nanowire samples.

\begin{figure}
\centering
\includegraphics[width=\linewidth]{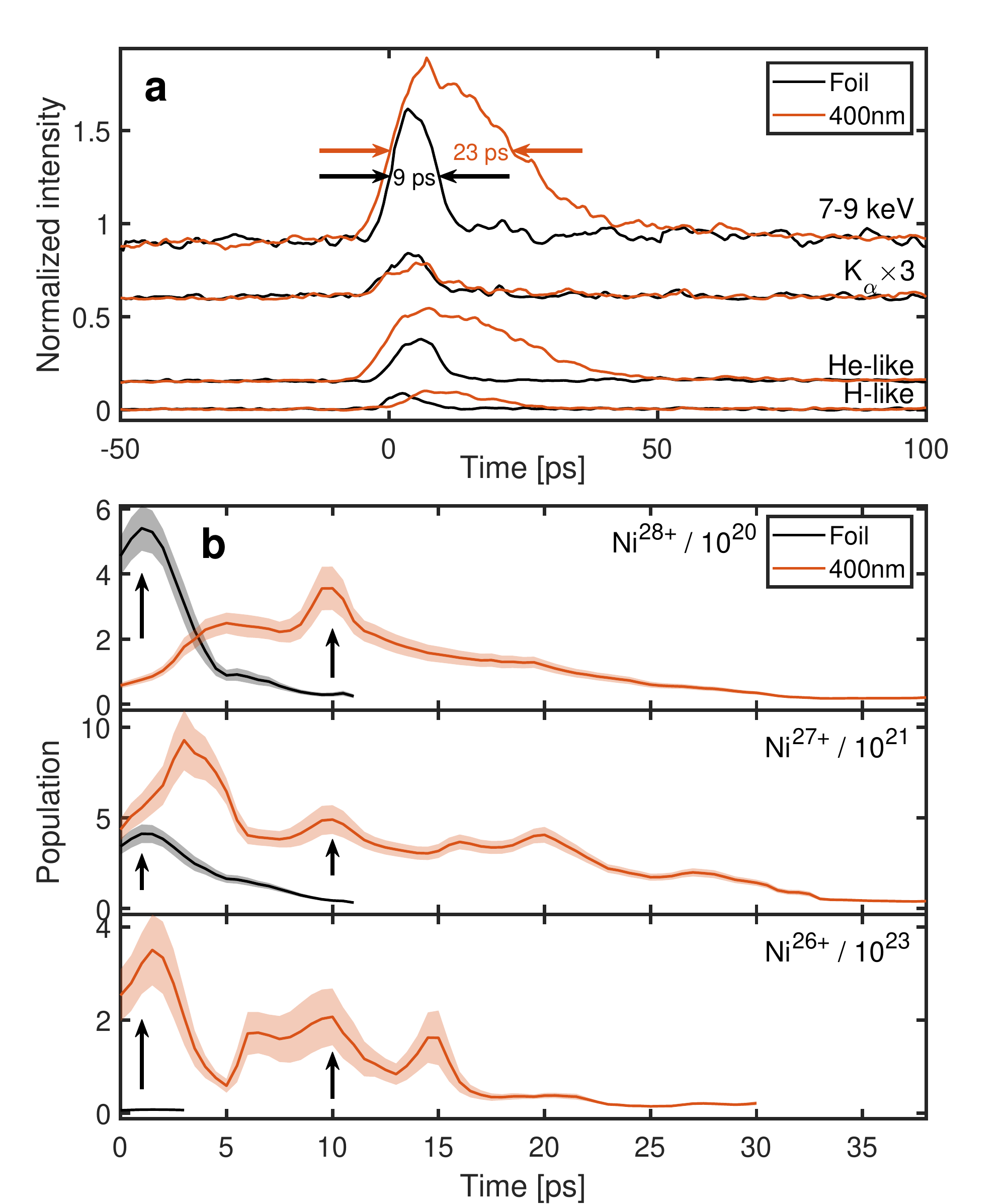}
\caption{a) Spectrally-integrated flat foil and \mbox{400 nm} wire array emission histories for the K$_{\alpha}$ (multiplied by three), H- and He-like complexes, normalized to the peak of the \mbox{7-9 keV} spectrally-integrated brightness, offset from each other for clarity. b) Ion populations of Ni$^{26-28+}$ obtained from the spectral modeling matched to the experimental data. Shaded regions denote the 1$\upsigma$ uncertainty in the fit caused by random noise in the streaked spectra. The arrows at early times indicate the production of highly ionized states by laser heating and at later times by inter-wire collisions. Note the different scales on the vertical axes.}
 \label{fig:flychk-populations}
\end{figure}

Immediately apparent from the streaked spectra in Figures \ref{fig:one} and \ref{fig:flychk-populations} is the enhancement in brightness of K-shell line emission, with the bulk of the spectrally integrated signal coming from helium-like nickel. Hydrogen-like emission is also enhanced. Both emit for around three times longer (FWHM) than in the case of a flat target, comparable to previous observations by Kulcs\'ar \emph{et al.} \cite{Kulcsar2000}.

\begin{figure}
\centering
\includegraphics[width=0.49\linewidth]{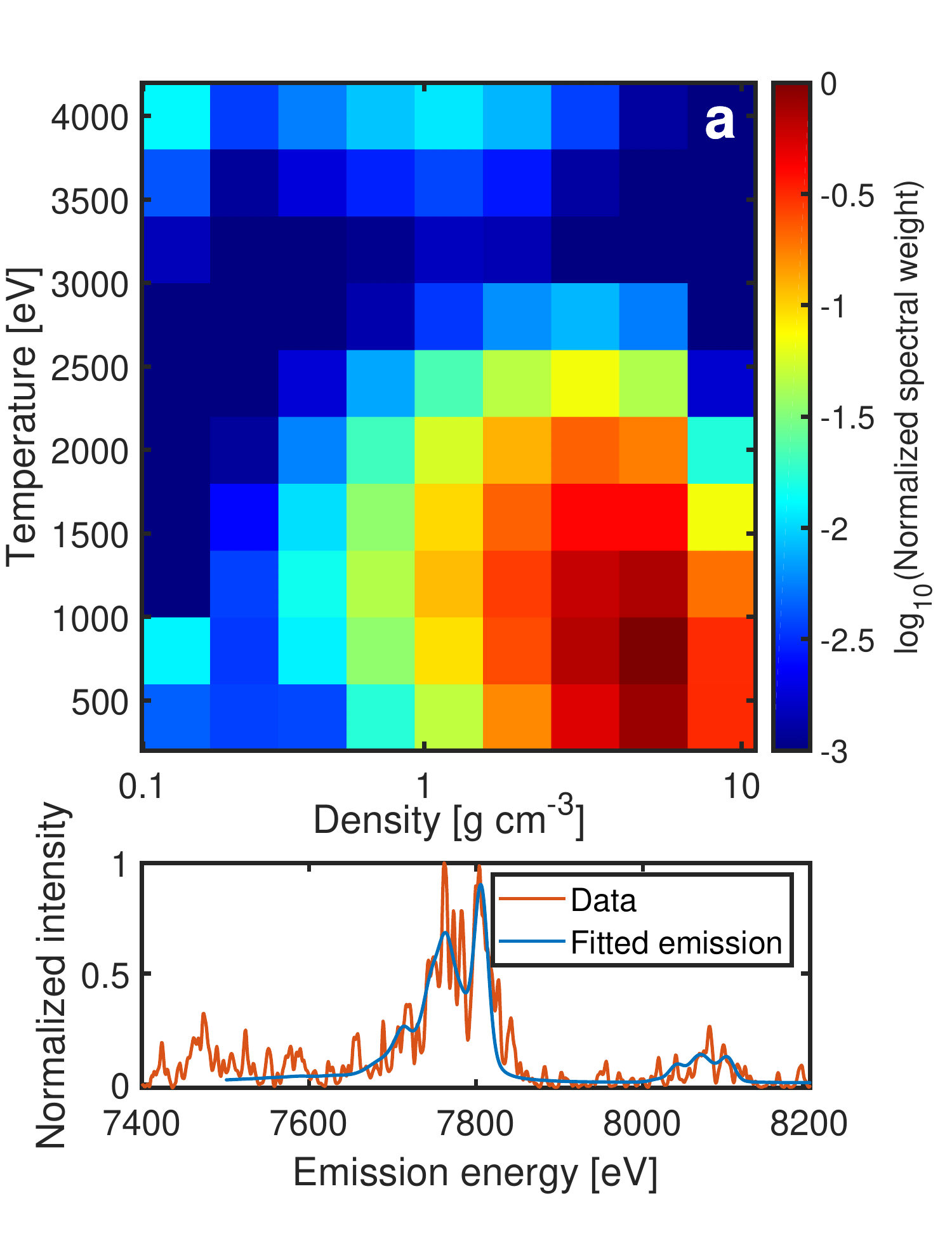}
\includegraphics[width=0.49\linewidth]{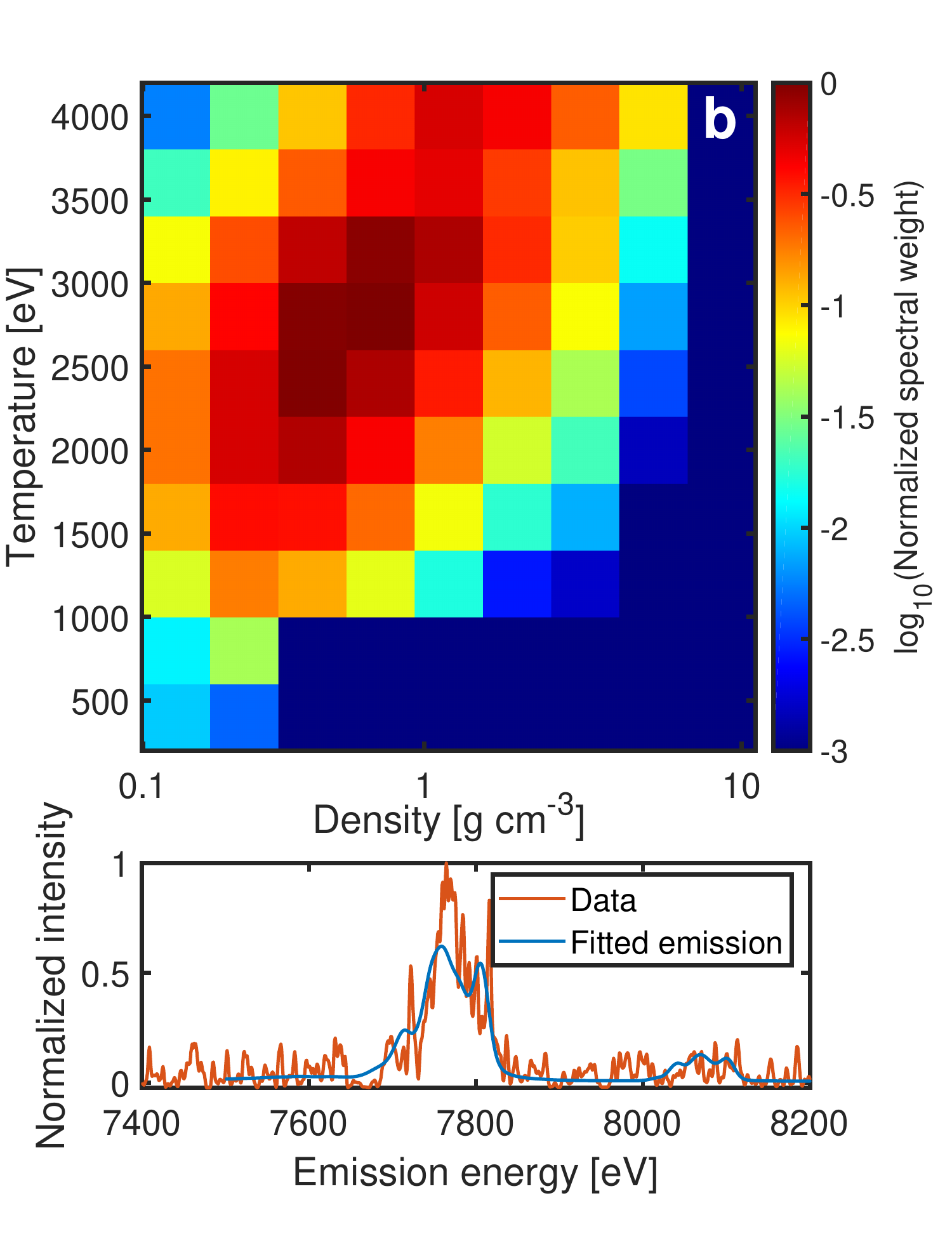}
\caption{Result of fitting the nanowire emission spectrum with FLYCHK using multiple temperatures and densities at (a)~2.5~ps and (b)~25~ps from the start of emission. Mapping conditions in this way shows a clear evolution in the plasma from initial laser heating through the subsequent wire explosions and re-heating from inter-wire collisions.}
 \label{fig:flychk_grid}
\end{figure}

\begin{figure*}
\centering
\includegraphics[width=\linewidth]{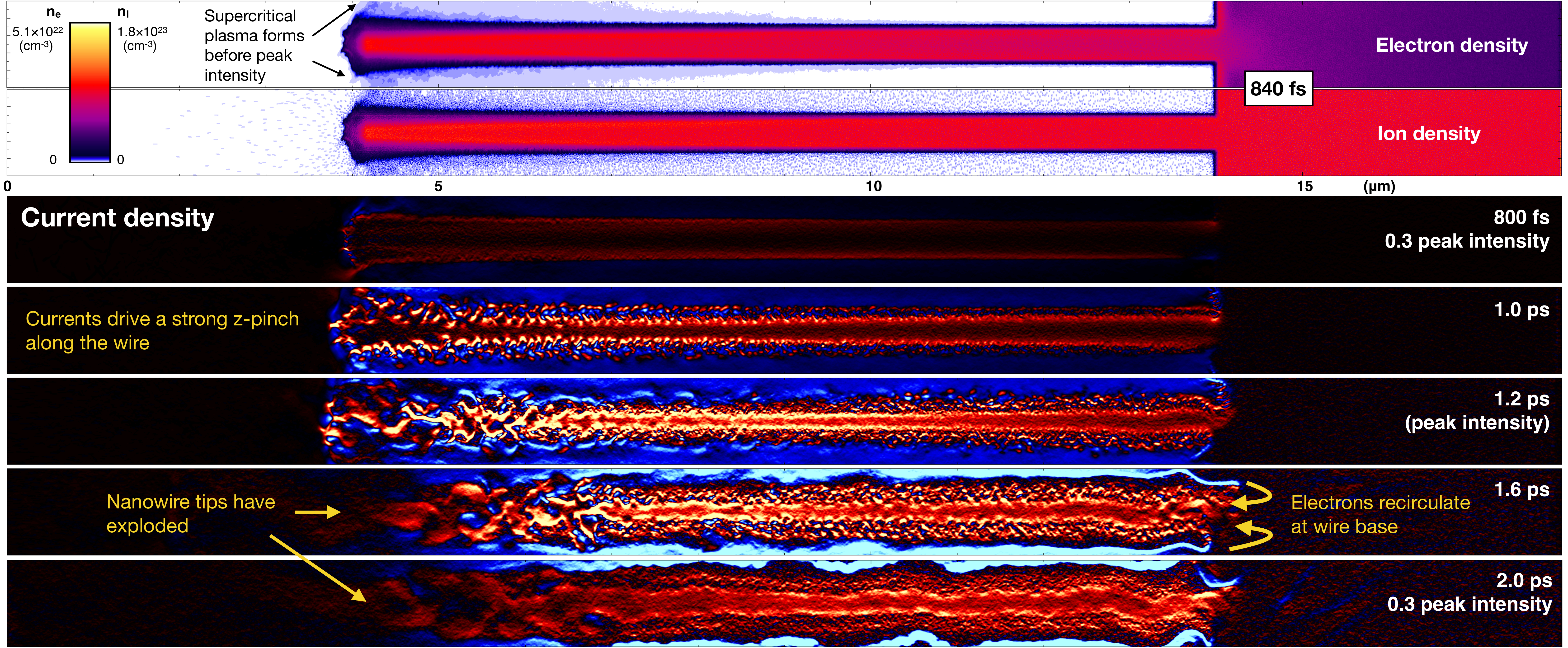}
\caption{2D PIC simulations showing the electron and ion density (top) and electron current (bottom) as the p-polarized gaussian \mbox{600 fs} laser pulse (incident from left) ramps up; the peak irradiance at the wire tip is at \mbox{t = 1200 fs}. Blue colors denote electrons moving to the right of the figure, red colors denote electrons moving to the left.}
\label{fig:pic}
\end{figure*}

Atomic kinetics simulations can be used to generate synthetic spectra to fit against the streaked emission, determining the evolution of the plasma conditions and ion populations \cite{Hoarty2013}. However, because the emission comes from a plasma containing gradients in both temperature and density that evolve over the observation window of $\sim$100~ps, no single condition well describes the emission profile for any given timestep. To address this issue we construct a temperature-density grid covering a full range of possible plasma conditions in the target, spanning temperatures from 0 to 4~keV and densities between 0.1 and 10~g/cm$^{3}$. For each point on this grid we perform a steady-state FLYCHK~\cite{chung2005flychk} simulation to calculate the corresponding X-ray emission spectrum and ion populations. A temperature-density distribution over this range of conditions is then used to construct a spectral fit to the data, with the shape of the distribution modeled by a set of free parameters. The number of free parameters is constrained by imposing an interpolation regime across parameter space, identified using scaling laws from an approximate focal spot profile, as determined experimentally via the KBXRM. The limited distribution shape allows a reduced number of free parameters to converge on a well conditioned fit to the data over the full temperature-density parameter space. 

We show the results of this process for two specific times, at 2.5~ps and 25~ps, in Fig.~\ref{fig:flychk_grid}. This fitting procedure is carried out for each time step, and allows us to extract average temperature and density conditions of the nanowire system as the plasma evolves. We see that at early times the system has an average temperature of around 1~keV and a density around 5~g/cm$^{3}$, while at later times the density decreases below 1~g/cm$^{3}$ and the temperature rises above 3~keV as the wires explode. From the FLYCHK simulations we can further extract the time evolution of the ion populations which we show in Fig.~\ref{fig:flychk-populations}. To estimate the uncertainty of this approach, an ensemble of temperature-density profiles was constructed, sampling different temperature-density points, allowing the deviation of the fit due to the limitations of both the signal and the imposed plasma parameter distribution function to be characterized with the variance in the outcomes of the ensemble. The 1$\upsigma$ uncertainties are shown as a shaded band in Fig.~\ref{fig:flychk-populations}. It was found that the signal to noise ratio on hydrogenic lines with one spectator electron diminishes significantly at later times - in particular for the case of the flat foil - limiting the time frame where the He-like ion population is well constrained.

Because our atomic kinetics modeling includes all charge states we can deduce the fully-ionized (Ni$^{28+}$) populations from the spectroscopic emission measurements even though these ions do not emit X-ray radiation. For the flat foil target both the fully ionized and the hydrogenic (Ni$^{27+}$) ion populations show an early peak in their evolution, as expected, on the order of the \mbox{2 ps} temporal resolution of the camera. The nanowires display a more complicated evolution leading to a peak in fully-stripped ions at around \mbox{10 ps}, well after the \mbox{0.5 ps}-duration laser pulse has turned off. The best fits to the spectra at these late times indicate sustained ion temperatures in excess of \mbox{1 keV} which would require a significant heating process not driven directly by the laser. This persistence of highly ionized states without an accompanying increase in K$_\upalpha$ emission can be explained by inter-wire collisions, where the short-pulse irradiated wires rapidly expand and radiate before the plasma flows from neighbouring wires collisionally re-heat and increase in ionization. 

Understanding the wire heating process requires 2D particle-in-cell (PIC) simulations covering the full duration of the laser pulse, performed using the PICLS code \cite{sentoku2008numerical} (Fig. \ref{fig:pic}). Periodic boundary conditions are used in the direction parallel to the nanowires to simulate neighbouring wires, and escaping boundaries otherwise. The grid resolution for the simulations is 5~nm/cell, with 8 ions per cell, and we include binary collisions, collisional impact ionization and field ionization in the calculation.

As predicted by Cristoforetti \emph{et al.}~\cite{cristoforetti2017transition}, the simulations show the formation of a thin critical-density plasma between the wire tips hundreds of fs before peak intensity is reached, even with no pre-pulse. The peak laser intensity does not penetrate between the wires as reported for fs-duration pulses~\cite{Purvis2013,kaymak2016nanoscale}, limiting the contribution of a Coulomb explosion to the wire dynamics. These are instead dominated by fast electron currents driven down the low-density regions between the wires before returning through the remaining solid wire core. By this mechanism the PIC simulations indicate initial heating of the solid wires above \mbox{2 keV} to a depth of \mbox{$>5$\textmu m} from the wire tips. The total heated volume is thus significantly enhanced compared with flat foils, where the heating depth remains very limited. The confinement of the majority of the energetic electrons to this hot plasma volume limits the neutral nickel population available to generate K$_\upalpha$, explaining why that signal does not deviate greatly from the unstructured foil target in intensity or in duration.

Despite the inability of the pulse to penetrate between the wires, $55-65\%$ absorption is predicted by the PIC code compared to $40\%$ for a flat target. The enhancement will be highest in the lower-intensity wings of the laser focal spot where the wire tip explosion and critical-density surface formation is slower, but is still present at peak intensity.

Further, the modeling suggests that transient mega-amp electron currents circulating through the wires generate powerful magnetic fields exceeding \mbox{$10^4$ T}, causing an axial pinch which pushes the wires briefly to 20\% above solid density in the core. The multi-Gbar pressures at these conditions drive a strong hydrodynamic explosion, radiating strongly as the wires expand towards each other through the low-density inter-wire plasma. Figure \ref{fig:one} shows that in 3D there exists a range of inter-wire separations in the sample; the 2D simulation parameters were chosen to represent the average of that range. Extending the simulations to 3D with sufficient resolution to capture the relevant physics over multiple picoseconds is challenging and the subject of ongoing efforts.

Scaling the performance of nanowire targets to kilojoule-scale lasers therefore requires consideration of the desired application. Highly relativistic interactions at \mbox{$>10^{20}$ W/cm$^{2}$} utilizing shorter pulses will drive a stronger pinch in the wires and a hotter velocity distribution of ions and electrons; desirable, for example, for driving fusion processes \cite{curtis2018micro}. The trade-off is against absorption efficiency since the inter-wire gap closure at high intensities is extremely fast. If maximum X-ray emission brightness is desired it may therefore be preferable to spread the available energy over a greater area of wires to create a larger, more uniform plasma volume, tuning the target and laser parameters to optimize the ion population for the desired emission lines. The potential parameter space is large and other factors such as laser pre-pulse and focal spot profile will likely mean that the optimum solution for each application will be facility-specific. For longer pulses in particular, wider-spaced structures are likely to permit greater inter-wire penetration and more efficient absorption, in exchange for a reduced number of emitters on the surface of a given target area. While the creation of homogeneous UHED states of matter require some amount of hydrodynamical equilibration, we note that based on the PIC results the early phase of wire evolution may prove to be an exciting platform for UHED physics experiments in its own right~\cite{kaymak2016nanoscale}.

The 2$\times$ enhanced brightness of the helium-like emission lines from flat to nanowire targets (around 3$\times$ when time-integrated, as shown in Figure \ref{fig:two}) and the prolonged duration of emission (from \mbox{9 ps} to \mbox{24 ps}, FWHM) is a direct demonstration of their potential as bright X-ray backlighters and invites further study of the internal hydrodynamic evolution of the sample. Enhanced X-ray sources such as these will be especially useful for diagnosing transient high-density plasmas only previously diagnosed using emission spectra \cite{Hoarty2013} and at facilities such as Laser M\'egajoule (LMJ), OMEGA and the National Ignition Facility (NIF) where kilojoule-scale short-pulse laser systems can be used as diagnostic drivers on a range of relatively long-duration experiments.

In conclusion we have presented picosecond resolution X-ray emission spectra from a nanostructured target driven at petawatt powers, showing a significant enhancement of K-shell -- in particular helium-like -- X-ray emission. 2D collisional PIC simulations attribute this primarily to improved laser coupling to the target, particularly in the lower-intensity wings of the laser spot. Analysis of the time-resolved emission spectroscopy signature suggests keV temperatures can be created volumetrically in several-micron-scale samples at near-solid density, making use of higher laser energies and pulse contrasts than previously studied. Our results show that significant populations of highly ionized states are generated over a significantly larger volume than in a flat target, suggesting several potential applications including short-pulse, high-fluence neutron sources, intense X-ray backlighters and UHED plasma sample generation. 

O.H., R.R., A.M. and S.M.V. acknowledge support from the U.K. EPSRC (EP/P015794/1), the Oxford University Centre for High Energy Density Physics (OxCHEDS) and the Royal Society. S.M.V. is a Royal Society University Research Fellow. The work of J.J.R., S.S., R.H., M.G.C., V.K. and A.P. was supported by the Fusion Sciences program of the Office of Science of the U.S Department of Energy Grant DE-SC0014610. The work of J.P., R.L. and R.T. was performed under the auspices of the US DOE by LLNL under contract no. DE-AC52-07NA27344.
\newline Data are \copyright~British Crown Owned Copyright 2020/AWE

\bibliography{hill-arxiv-nanowires}

\renewcommand{\theequation}{A\arabic{equation}}
\setcounter{equation}{0}

\end{document}